# The Markup Language for Designing Gaze Controlled Applications

J. Matulewski[1], B. Bałaj[1], I. Mościchowska[2], A. Ignaczewska[1], R. Linowiecki[1], J. Dreszer[1], W. Duch[1]

[1]*Nicolaus Copernicus University in Toruń, Poland*

[2]*AGH University of Science and Technology in Cracow, Poland*

**Authors' Mini-bios:**

**Jacek Matulewski** (jacek@phys.uni.torun.pl) is a physicist and programmer with an interest in quantum optics and eyetracking; he is an assistant professor in the Department of Physics, Astronomy and Informatics of Nicolaus Copernicus University (NCU) in Toruń, Poland and the leader of the Therapeutic Games Laboratory *GameLab* at NCU Centre for Modern Interdisciplinary Technologies.

**Bibianna Bałaj** is a psychologist with an interest in cognitive psychology and eyetracking; she is a researcher in the Centre for Modern Interdisciplinary Technologies at NCU.

**Iga Mościchowska** is a practitioner in interaction design and user research, with academic background in sociology; she is affiliated with AGH University of Science and Technology in Cracow, Poland, where she is a head of a postgraduate course in User Experience & Product Design.

**Agnieszka Ignaczewska** is a scientific assistant, specializing in social communication science from a cognitive perspective. A graduate of the NCU. She worked at the Centre for Modern Interdisciplinary Technologies at NCU as an assistant during implementation of the scientific project entitled "NeuroPerKog: Phonematic hearing tests and working memory of infants and children". She currently works at a research lab in the NCU Faculty of Humanities, as an assistant in research projects.

**Rafał Linowiecki** is a student of informatics at the Department of Physics, Astronomy and Informatics of NCU. Part of the software presented in the paper was developed for his diploma thesis.

**Joanna Dreszer** is a psychologist with an interest in cognitive psychology and individual differences; she is an assistant professor in the Department of Psychology, Faculty of Humanities, NCU.

**Włodzisław Duch** is the head of the Neurocognitive Laboratory in the Centre for Modern Interdisciplinary Technologies, and for many years has been in charge of the Department of Informatics at NCU. In 2014-15 he served as a deputy minister for science and higher education in Poland, and in 2011-14 as the Vice-President for Research and ICT Infrastructure at his University. Before that he worked as a Nanyang Visiting Professor (2010-12) in the School of Computer Engineering, Nanyang Technological University, Singapore, where he also worked as a visiting professor in 2003-07. He has also worked at the University of Florida; Max-Planck-Institute, Munich, Germany; Kyushu Institute of Technology, Meiji and Rikkyo University in Japan; and several other institutions. He has been a member of the editorial board of IEEE TNN, CPC, NIP-LR, Journal of Mind and Behavior, and 14 other journals. He was a co-founder & scientific editor of the Polish Cognitive Science journal, has served two terms as the President of the European Neural Networks Society executive committee (2006-2008-2011), and is an active member of IEEE CIS Technical committee. The Board of Governors of the International Neural Network Society elected him for their most prestigious College of Fellows.

**Notes**

*Acknowledgments*

We would like to express our gratitude to Magdalena Szmytke, Jan Nikadon and Piotr Nikadon for revision of English translation of GIML keywords, to Nicolas Jędrzejczak-Rey and Jarosław Zaremba for French translation, to Miłosz Kłyszewski, Joanna Pogodzińska, Joanna M. Pogodzińska and Sławomir Pogodziński for German translation and to Anita Pacholik-Żuromska for its revision. We would also like to mention the contribution by Jakub Szwarc, who prepared the set of games using GIML, as a part of his diploma thesis, and tested the language and its interpreter.

*Funding*

The GIML project is part of the grant *NeuroPerKog: development of hearing and working memory in infants and children* funded by Polish National Science Centre (UMO-2013/08/W/HS6/0033).

# CONTENTS

# 1. MOTIVATION

Software development is time consuming and requires expert knowledge, therefore it is usually very expensive, especially in the case of specialized applications, as for example those that use eye trackers and other gaze tracing devices. On the other hand users of this particular type of applications, that is individuals with motor disabilities and generally with limited communication skills (e.g. resulting from stroke or traffic accidents), are very often in difficult financial situation. Therefore there is a need for a new technique of creating gaze controlled software which is relatively cheap and simultaneously easy to use by non-programmers. Thus, we wanted to design a human-readable language dedicated for this purpose, and to prepare its interpreter. We called this language GIML, which stands for *Gaze Interaction Markup Language* and its interpreter is GCAF, which stands for *Gaze Controlled Application Framework*.

Another use that we had in mind while designing GIML is related to preparation and performance of psychological experiments involving gaze interaction between a subject and a computer system. Both GIML and GCAF have already been used for several years in experiments performed with infants. The advantages of GIML in the process of designing psychological experiments lie in its full interactivity and flexibility. The order of graphical display objects (the experiment path) can be selected by the subject, but it can also be set, or narrowed down, by the investigator. The latter can also determine the conditions under which the stimuli are visible to the subject. As a result it is possible to design experiments that can be controlled both by a person carrying the experiment and by study participants, depending on what is needed.

## 2. EYETRACKING AND GAZE INTERACTION

The common saying "eyes are the only moveable part of the brain" clearly expresses the idea of gaze interaction as a method of communication used both in natural social interactions, starting from infant-parent communication (Fogel, Nwokah, Hsu, Dedo, & Walker, 1993), and by people with limited capacities for using common communication methods, e.g. those with physical disabilities, cerebral palsy (Galante & Menezes, 2012), as well as young children (Borgestig et al., 2017). We often communicate our intentions using eye movements, e.g. by gazing at a person we want to talk to in a group of people (*diaxis*) (Smith, Vertegaal, & Sohn, 2005). We look before we act, sometimes without a conscious thought. Our behavior is also guided by gaze (Hayhoe & Ballard, 2014). Various low-tech systems are available for people with communication difficulties (see: Zupan & Jenko, 2012), e.g. communication boards (Zawadzka & Murawski 2015). Computer-based gaze interaction systems, which use eye tracking technology, are the main tools for augmentative and alternative communication (AAC) technology (Clarke & Bloch, 2013) and allow patients to make autonomous decisions when and how to express their needs and begin to communicate with others (for review see: Nerișanu, Nerișanu, Maniu, & Neamțu, 2017). People can write messages with their eyes using a number of available applications (Majaranta & Räihä, 2002; Mott, Williams, Wobbrock, & Morris, 2017), for example *Dasher* (Rough, Vertanen, & Kristensson, 2014), *GazeTalk* (Hansen 2001), *pEye* (Huckauf & Urbina, 2008), *Eye-S* (Porta & Turina, 2008), *Quickwriting* (Perlin, 1998; Bee & Andree, 2008) or *Stagazer* (Hansen et al. 2008).

In addition to the above, many people like to spend their leisure time performing computer activities (e.g. gaming or social media). There are several games developed especially for gaze interaction, e.g. Eagle Aliens (Betke, 2010), but developers often try to enable gaze-interaction in existing games, as well (see: Istance, 2017). The *Tobii Game Hub* package also enables eye tracking and head tracking in a rather large group of games, which are listed at Tobii's web page (Tobii, 2019). A system for control of on-line video playback was developed in our lab (Matulewski et al., 2018). Notably, Windows 10 operating system (from December 2018) includes built-in eye tracking support for Windows Store Apps, which co-operates with Tobii eye trackers. However, gaze-interaction can be used in far more ways: it enables environmental control (for review see: Majaranta, et al., 2012), driving a wheelchair (Wästlund, Sponseller, & Pettersson, 2010), and control of home appliances (Bonino, Castellina, Corno, De Russis, 2011). Therefore the use of eye trackers can re-establish some level of autonomy for disabled people (Calvo et al., 2008). Communication and control improves their quality of life (Majaranta, et al., 2012) preventing depressive states which often affect patients with spinal cord injuries. Gaze interaction technology improves social integration of cerebral palsy patients, their devotion to the training, leading to decreasing level of abandonment (Nerișanu, Nerișanu, Maniu, & Neamțu, 2017).

## 3. EXPLANATION OF THE DESIGN CHOICES

As mentioned above, we propose new programming language, GIML, and provide its interpreter, which allows to easily create gaze controlled applications, even by non-programmers. To fulfill this requirement, GIML should be intuitive - its keywords should be taken from the user's native language and the code design should ultimately be supported by some kind of debugger helping to find errors. In fact, our first thought was not to create the programming language, but a specialized visual editor for designing application content. But we soon realized that such editor, even in the present state of GIML development, would have so many features and options that it would make it very complicated and unfriendly, with a rather steep learning curve. The code writing instead seems to be a much more flexible solution, although it may be frightening for beginners. Importantly, a very small subset of keywords is sufficient to write a simple application. This is quite similar as in the case of HTML (*Hyper-Text Markup Language*): although visual editors are available, and are now already quite handy, most experienced programmers edit directly the HTML code. It is a faster and ultimately also an easier way of creating and modifying web pages. It is the same in the case of XAML (*eXtensible Application Markup Language*) – the language designed by Microsoft for description of interfaces in advanced applications for Windows. Like these two languages, GIML is based on the XML text file format, therefore it belongs to the group of the so-called markup languages. Thus one can write the code using any editor, even the simple notepad. GCAF provides support for spelling and syntax checker. However, at the present stage it is not integrated with any editor.

The choice of a markup language as the base for GIML is important from the viewpoint of its intended accessibility for non-programmers. Beginners can easily learn the main elements of GIML – there are only a few of them. Also only a few attributes are indispensable to write simple applications. However, the full list of attributes is quite long, but they can be learned gradually as they are needed.

The structure of the GIML code (i.e. the hierarchy of XML elements) reflects the structure of the set of scenes saved in a single file: the `scenes` element is the container for the set of `scene` elements, which in turn contain regions represented by the elements `region`. Regions have two functions: they provide places for displaying texts, images, animated images or movies (the sounds can also be attributed to them), but at the same time regions are subject of gaze interaction and can change their state depending on the user's visual attention. A region can be in one of three states (cf. Fig. 1): 1) non-active, when the user is not looking at it, 2) activated, if the user takes a glance at this region, and finally 3) reacting, when the user is staring at the region for longer time (by default longer than one second, but this can be adjusted). The two latter states are defined by dedicated sub-elements of `region`, namely `activation` and `reaction`, which are the lowest level elements of the GIML structure (cf. Fig. 2). The details of all the states of regions are provided using attributes and will be partly described in the following sections.

**Figure 1. States of regions and possible transitions between them**

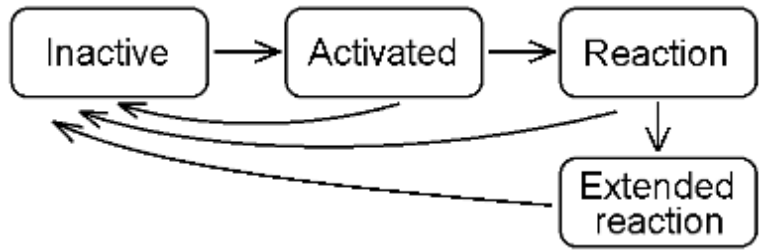

**Figure 2. Scheme of the part of the GIML code responsible for scene description**

```
<Settings ...>
    <Scenes ...>
        <Scene ...>
            <Region ... >
                <Activation ... />
                <Reaction ... />
            </Region>
            ...
        </Scene>
        ...
    </Scenes>
</Settings>
```

GIML allows the creator of application to design the way the user can navigate between the scenes and the way she or he can manage the visibility of the regions on the specific scene. However, both features are optional and the application (or experiment) can also have a fixed storyboard. Region managing includes enabling and disabling the regions (i.e. switching them on and off). It is done by attributing actions to the regions which play the role of triggers. The concept of the scenes and regions and their dynamics are the key feature of GIML, which makes this language highly flexible. It allows to design and create many types of applications, including interactive experiments, personalized communication boards or even gaze-controlled games.

Regions can display images and movies, different in various regions depending on their states. In addition sounds can be associated with any state of the region. In order to avoid using file paths of these resources directly inside the scene codes, which is difficult to maintain, we decided to use a two-level structure in this case: all resources have to be declared, which is done by placing them in one of three resource dictionaries, separately for images, sounds and movies, then they can be referred from the scene codes by the declared names. This approach additionally allows to eliminate multiple memory allocations for the same resource. More advanced designers may use the idea of lists of resources or their attribute values (e.g. colors). This allows one to choose several scenarios for switching the stimuli as described below.

In early versions of GIML, the `style` element was implemented; it allowed to define the set of attribute values, which could be set to regions (a similar solution of styles is used in XAML or CSS in HTML). However, this concept was very difficult to understand for non-programmers and also was not really useful, thus it was removed from the language specification. Instead we have introduced the so-called templating, which is similar to class inheritance in algorithmic languages. One can define the template scenes containing some regions and settings. Other scenes can use these templates and then the regions defined in the template appear on the screen together with the ones defined in the inherited scene.

## 4. GAZE INTERACTION MARKUP LANGUAGE

In this section version 1.3 of the *Gaze Interaction Markup Language* specification is described, as it has already been implemented in our language interpreter, called *Gaze Controlled Application Framework* (GCAF). The newly proposed language elements designed for text entry, elements supporting reading examinations or those for offset correction during the runtime and support for blinks based application control are skipped here, since their implementation and testing is still in progress.

GIML is a declarative language, it describes a final effect (e.g. the appearance and behavior of application in the interaction with the user) and not the algorithm that leads to such effect. This makes it easier to learn and use for people who are not professional programmers. As already mentioned, GIML is based on the XML file format. Its full tree-like structure of elements is visible in Fig. 3. Most of the elements, especially the lower level ones, can have many attributes that are listed at our web page (Matulewski, 2019).

**Figure 3. The GIML elements structure**

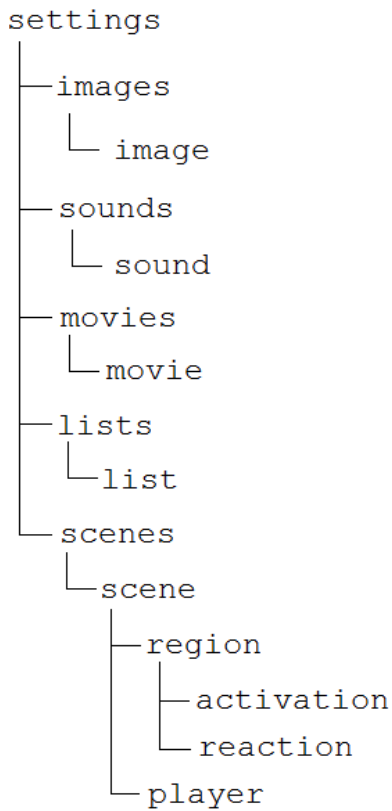

GIML is case insensitive – it is because the case sensitiveness caused many errors in pilot studies we performed. The only exceptions are made for the values of the `text` attribute of the `region` element and values of the `value` attribute in lists, where the original spelling from the file is preserved and displayed on screen.

At the moment four GIML language versions are available and supported by the current edition of the GCAF framework: English, German, French and Polish. Also a translation tool is available; it allows to easily translate any GIML file from or into all those languages.

**Resources**

All the resources have to be declared in one of the three containers: `images`, `sounds` or `movies`. The declaration of resource provides an opportunity to set the volume of sounds and movies, define the number of repetitions, or specify transparent color in the case of images (see details in Matulewski, 2019). In the resource declaration one can use both the absolute or relative path to the resource file. In the latter case, the folders given by the value of the `folder` attribute of the root element and the resource parent element (namely the `settings` and `images` elements) may be used as a reference folder. If it is not present, the working directory of GCAF is assumed.

**Figure 4. An example of resource declaration**
```
<settings folder="C:\Users\Jacek\GIML\Assets" language="en">
    <images folder="img">
        <image name="img1" path="img1.png" />
    </images>
    ...
</settings>
```

Let us consider the image resource declaration as an example. The simplest form of such an element along with its surrounding elements is shown in Fig. 4. The `image` element contains two attributes: `name` and `path`. Both are required. The first one specifies the name of the image, by which it will be referred from the scene code. The second one points to the file by its relative path. Taking into account the resource folder declared in the `settings` element, and the subfolder declared in the `images` container, the full path of the file in this example is *C:\Users\Jacek\GIML\Assets\img\img1.png*. It is possible to extend the specification of the image element by adding some additional optional attributes. One example is `transparencyKey`, the value of which specifies color (given by its name or by the HEX string). It points to the color which will be changed to the transparent one while the scene is being rendered on the screen. The other `image` element attributes are used only for the animated images (see Matulewski, 2019).

**Regions and their states**

Scenes are constructed using the `scenes` elements, which have three obligatory attributes, i.e. the name of the default scene, as well as the width and height of the screen.

Using additional attributes it is also possible to declare the scene to which one navigates after pressing the *Pause* key on the keyboard and to enable the "flashlight" effect limiting the visible range of scene to a given radius around the gaze position (in order to limit the peripheral vision). In Fig. 5 the `scenes` container and one `scene` named *scene1* is defined. The name is the only mandatory attribute of the `scene` element. This scene is defined as a default one. An optional attribute of the `scene` element, namely the `backgroundColor` indicating the scene's background color, is also used. Other attributes that may appear here include: the background image and sound, attributes controlling the optional blackout of scene if one region is selected, the flag deciding whether the state of the scene is reset after it is revisited from another scene, and many others.

**Figure 5. An example of scene definition**
```
<settings folder="C:\Users\Jacek\GIML\Assets" language="en">
    <images folder="img">
        <image name="img1" path="img1.png" />
    </images>
    <scenes nameOfDefaultScene="scene1"
            originalScreenSizeX="1024" originalScreenSizeY="768">
        <scene name="scene1" backgroundColor="beige">
        </scene>
    </scenes>
</settings>
```

As mentioned before, the scene is the container for the regions. The `region` element is used most frequently. It has several obligatory attributes determining its name, shape, location and size. Any other attributes are optional (see. Matulewski, 2019 for the list). One of the mandatory attributes is the `nameOfImage`, which points to the image declaration in the resource dictionary. Fig. 6 shows an example of a simple region located at the top-left corner of the screen presenting the image declared in Fig. 4.

**Figure 6. An example of region definition**
```
<settings folder="C:\Users\Jacek\GIML\Assets" language="en">
    <images folder="img">
        <image name="img1" path="img1.png" />
    </images>
    <scenes nameOfDefaultScene="scene1"
            originalScreenSizeX="1024" originalScreenSizeY="768">
        <scene name="scene1" backgroundColor="beige">
            <region name="region1" shape="rectangle"
                    locationOfCenterX="300" locationOfCenterY="200"
                    sizeX="200" sizeY="200"
                    nameOfImage="img1" />
        </scene>
    </scenes>
</settings>
```

One can specify the size and location of a region using the absolute values given in pixels. However, it can be inconvenient if various screen sizes are used. Therefore it is suggested that relative values should be used. Fig. 7 presents a modified code of the region from Fig. 6 in which the location and size was set as a percent value of the size of the screen (the same approach is also used in HTML). This way of providing values can be also used in the case of other region attributes.

**Figure 7. The use of relative values for determining the location and size of regions**
```
<settings folder="C:\Users\Jacek\GIML\Assets" language="en">
    <images folder="img">
        <image name="img1" path="img1.png" />
    </images>
    <scenes nameOfDefaultScene="scene1"
            originalScreenSizeX="1024" originalScreenSizeY="768">
        <scene name="scene1" backgroundColor="beige">
            <region name="region1" shape="rectangle"
                    locationOfCenterX="30%" locationOfCenterY="20%"
                    sizeX="20%" sizeY="20%"
                    nameOfImage="img1" />
        </scene>
    </scenes>
</settings>
```

The region in Fig. 7 is not interacting with the user. It would not change its appearance no matter if the gaze position is located inside it or not. To achieve such effect one needs to define the `activation` and `reaction` sub-elements. Both have no obligatory attributes. Using these sub-elements one can change all the properties of the region, excluding its position and size (although translation of the region is possible as a result of activation or reaction). For example the displayed image can be changed as shown in Fig. 8.

**Figure 8. Using the region states for changing the displayed image**
```
<settings folder="C:\Users\Jacek\GIML\Assets" language="en">
    <images folder="img">
        <image name="img1" path="img1.png" />
        <image name="img2" path="img2.png" />
        <image name="img3" path="img3.png" />
    </images>
    <scenes nameOfDefaultScene="scene1"
            originalScreenSizeX="1024" originalScreenSizeY="768">
        <scene name="scene1" backgroundColor="beige">
            <region name="region1" shape="rectangle"
                    locationOfCenterX="300" locationOfCenterY="200"
                    sizeX="200" sizeY="200"
                    nameOfImage="img1">
                <activation nameOfImage="img2" />
                <reaction nameOfImage="img3" />
            </region>
        </scene>
    </scenes>
</settings>
```

Notably, the essential feature of GIML is that the attributes can determine not only the application appearance, but also its behavior, namely the behavior of the regions, including their responsiveness to the user's gaze. For all three states one can specify the text, font, color and thickness of the frame surrounding the region, as well as the image, sound or movie associated with this region. One can also specify animation of the region or animation of its contents. Regions can also be disabled, to make them invisible, and then re-enabled. Areas are enabled by default, but one can declare an area that is initially disabled and is enabled by activation or reaction of another region in the scene. This allows for a complete change of the scene content as a result of its interaction with the user.

Fig. 9 shows another example of a region that displays the label showing the name of each state. The color of text is also changed, but the font family and its size stays the same. To change this, the attributes of `font`, `fontSize`, `fontColor` and `fontStyle` have to be repeated in sub-elements describing the activation and reaction states. Otherwise the default values are used.

**Figure 9. Displaying and formatting the text in regions**

```
<region name="region1" shape="rectangle"
        locationOfCenterX="300" locationOfCenterY="200"
        sizeX="200" sizeY="200"
        text="normal state" fontColor="Black" fontSize="20">
    <activation text="activation state" fontColor="Brown" fontSize="20" />
    <reaction text="reaction state" fontColor="SandyBrown" fontSize="20" />
</region>
```

Another important attribute of the region is `tag`. It sets the messages sent to the eye tracker server which are included in the recorded raw-data files when the state of the area changes (supported for example by SMI eye trackers). This makes the oculographic data analysis much easier.

A return from the reaction to the normal state occurs by default when the gaze leaves the region. However, it is possible to extend the duration of the reaction state. There are two additional conditions: waiting for the end of the sound played in the reaction state or waiting until predefined interval of time elapses. This can be set using the region attribute `conditionOfReactionCompletion`, which may take three values: `RegionLeave`, `SoundEnding` and `TimeElapsed`. In the latter case additional attribute `reactionDuration` is used to set the number of milliseconds by which the reaction time will be extended (Fig. 10). Notably, in all the cases the primary condition of moving the gaze position out of the region must be fulfilled.

**Figure 10. Extending the reaction completion condition**

```
<region name="region1" shape="rectangle"
        locationOfCenterX="300" locationOfCenterY="200"
        sizeX="200" sizeY="200"
        nameOfImage="img1"
        conditionOfReactionCompletion="TimeElapsed" reactionDuration="5000">
    <activation nameOfImage="img2" />
    <reaction nameOfImage="img3" />
</region>
```

It is also possible to force a direct transition to the reaction state. In this case one needs to define the attribute `automaticReactionAfterTime` equal to the number of milliseconds after which the reaction state of the region automatically starts. This allows the researcher to impose a fixed transition time defining change of a scene.

**Scene navigation**

An alternative to disabling some regions and enabling others in order to change the content of the scene involves switching between the whole scenes. GIML allows to define multiple scenes in a single file, which is rather unusual compared to other similar languages (e.g. HTML or XAML). However, the navigation between scenes is one of the key features of our language. The scene switching may be set by changing the state of the region, typically to reaction state. Other possible actions include: drawing the "attention frame", moving the active area to another location and resetting the region or scene. Navigation between scenes becomes essential in the case of more complicated applications. By default the scene retains its state after it is left and it is restored upon return. However, it can be reset if the scene attribute `resetAfterEnter` is used. Fig. 11 shows the code of two scenes. The user can switch between them back and forth. Of course it is possible to define more scenes and the application designer can choose any path for switching among them or give the choice to the user.

**Figure 11. Example of two scenes and navigation between them**

```
<settings folder="C:\Users\Jacek\GIML\Assets" language="en">
    <images folder="img">
        <image name="img1" path="img1.png" />
        <image name="img2" path="img2.png" />
        <image name="img3" path="img3.png" />
    </images>
    <scenes nameOfDefaultScene="scene1"
            originalScreenSizeX="1024" originalScreenSizeY="768">
        <scene name="scene1" backgroundColor="beige">
            <region name="region1" shape="rectangle"
                    locationOfCenterX="300" locationOfCenterY="200"
                    sizeX="200" sizeY="200"
                    nameOfImage="img1">
                <activation nameOfImage="img2" />
                <reaction actionType="TransitionToScene"
                          nameOfTargetScene="scene2" />
            </region>
        </scene>
        <scene name="scene2" backgroundColor="black">
            <region name="region1" shape="rectangle"
                    locationOfCenterX="300" locationOfCenterY="200"
                    sizeX="200" sizeY="200"
                    text="Return" fontColor="Beige" fontSize="20">
                <activation text="Return"
                            fontColor="SandyBrown" fontSize="20" />
                <reaction action="TransitionToScene"
                          nameOfTargetScene="scene1" />
            </region>
        </scene>
    </scenes>
</settings>
```

Let us mention one of the many options important in the context of scene navigation, namely the possibility to suspend the scene switching until the end of the extended reaction of the region (e.g. if the action involves the region translation or sound playing). It is controlled by the attribute `hangSceneTransitionUntilReactionCompleted`. As mentioned before, the condition of finishing the reaction can be controlled by the attribute `conditionOfReactionCompletion`.

**Varying the stimulus: lists and randomization**

Two additional mechanisms allow to vary the content of the scenes: randomization and lists of values. The first one is rather simple. Instead of one value, one can provide a series of values separated by colons with the keyword `rand` at the front. For example the font color can be chosen randomly from seven colors given using the following attribute:

```
fontColor="rand:Red:Orange:Yellow:Green:Blue:Indigo:Violet"
```

In the case of numerical values, e.g. the position or size of the region, one may set up the ranges of random values:

```
locationOfCenterX="rand:200:400"
```

Random values are drawn only once during the initialization of application, so they vary between application runs, but are fixed for one run, even when many scenes are visited. Lists are the second mechanism allowing to vary the content of the GIML applications. This is important especially in the process of designing psychological experiments, in which usually many trials of the same task are presented and only some elements of the stimuli are changed. In such a case one can define a list of colors, positions or image names and change these sequentially or randomly (by drawing from the list with or without returns). Moreover one can control the moment when the specific scenes are switched and the subsequent values are chosen. Fig. 12 provides an example presenting a modification in the code from Fig. 11. Two lists, that is the elements `list` in container `lists`, are defined: one for images and one for colors. In the first one items are drawn with no returns (default option), while in the second the elements are taken sequentially, which is determined by the attribute `drawing`. Regions named *region1* in both scenes refer to the lists. The region from the scene *scene1* refers to the list of images in the attribute `nameOfImage`. Note that the symbol @ is used before the name of the list. Similarly the region *scene2* refers to the list of colors in the attribute `fontColor`. In addition, the scene *scene1* contains the attribute `nameOfListsSwitchedOverAfterEnter` with a value containing list names, which provide a signal to these lists to select a new value. Thus any entrance to the scene *scene2* causes a change of the image visible in the region *region1*. In turn, another scene attribute `nameOfRegionEnabledAfterListFinished` enables the region *region2* in this scene if any of the lists have no more elements to be drawn. As a result the text "End of lists" becomes visible after the fourth entry to the scene *scene1*.

**Figure 12. Example of using lists as a source of resources**
```
<settings folder="C:\Users\Jacek\GIML\Assets" language="en">
    <images folder="img">
        <image name="img1" path="img1.png" />
```

```xml
            <image name="img2" path="img2.png" />
            <image name="img3" path="img3.png" />
        </images>
        <lists>
            <list name="imgs" elementsType="Strings"
                values="img1;img2;img3" />
            <list name="colors" elementsType="Colors"
                values="Red;Orange;Yellow" drawing="Sequentialy" />
        </lists>
        <scenes nameOfDefaultScene="scene1"
                originalScreenSizeX="1024" originalScreenSizeY="768">
            <scene name="scene1" backgroundColor="beige"
                    nameOfListsSwitchedOverAfterEnter="imgs;colors"
                    nameOfRegionEnabledAfterListFinished="region2">
                <region name="region1" shape="rectangle"
                        locationOfCenterX="300" locationOfCenterY="200"
                        sizeX="200" sizeY="200"
                        nameOfImage="@imgs">
                    <reaction actionType="TransitionToScene"
                            nameOfTargetScene="scene2" />
                </region>
                <region name="region2" shape="rectangle" enabled="no"
                        locationOfCenterX="300" locationOfCenterY="600"
                        sizeX="200" sizeY="200"
                        text="End of list" fontColor="brown" fontSize="20" />
            </scene>
            <scene name="scene2" backgroundColor="black">
                <region name="region1" shape="rectangle"
                        locationOfCenterX="300" locationOfCenterY="200"
                        sizeX="200" sizeY="200"
                        text="Return" fontColor="@colors" fontSize="20">
                    <reaction actionType="TransitionToScene"
                            nameOfTargetScene="scene1" />
                </region>
            </scene>
        </scenes>

</settings>
```

One can gather the lists in groups, in which elements are selected synchronously. For example we can add a list of strings, which contains the labels for images and assigns these to the same group as the list of images (Fig. 13). This assures that the elements with the same index will be chosen from both lists and thus the label will fit the image.

### Figure 13. Example of grouping lists

```xml
<settings folder="C:\Users\Jacek\GIML\Assets" language="en">
    <images folder="img">
        <image name="img1" path="img1.png" />
        <image name="img2" path="img2.png" />
        <image name="img3" path="img3.png" />
    </images>
    <lists>
        <list name="imgs" elementsType="Strings"
            values="img1;img2;img3" group="1" />
    <list name="colors" elementsType="Colors"
        values="Red;Orange;Yellow" drawing="Sequentialy" />
    <list name="captions" elementsType="Strings"
        values="img1;img2;img3" group="1" />
    </lists>
    <scenes nameOfDefaultScene="scene1"
            originalScreenSizeX="1024" originalScreenSizeY="768">
```

```
            <scene name="scene1" backgroundColor="beige"
                   nameOfListsSwitchedOverAfterEnter="imgs;colors;captions"
                   nameOfRegionEnabledAfterListFinished="region2">
                <region name="region1" shape="rectangle"
                        locationOfCenterX="300" locationOfCenterY="200"
                        sizeX="200" sizeY="200"
                        nameOfImage="@imgs">
                    <reaction actionType="TransitionToScene"
                              nameOfTargetScene="scene2" />
                </region>
                <region name="region2" shape="rectangle" enabled="no"
                        locationOfCenterX="300" locationOfCenterY="600"
                        sizeX="200" sizeY="200"
                        text="End of list" fontColor="brown" fontSize="20" />
                <region name="region3" shape="rectangle"
                        locationOfCenterX="300" locationOfCenterY="500"
                        sizeX="200" sizeY="200"
                        text="@captions" fontColor="brown" fontSize="20" />
            </scene>
            <scene name="scene2" backgroundColor="black">
                <region name="region1" shape="rectangle"
                        locationOfCenterX="300" locationOfCenterY="200"
                        sizeX="200" sizeY="200"
                        text="Return" fontColor="@colors" fontSize="20">
                    <reaction actionType="TransitionToScene"
                              nameOfTargetScene="scene1" />
                </region>
            </scene>
        </scenes>
</settings>
```

Notably, in this example a separate region is used for image captions. We could use the same region as well, shifting the image or text. This can be done using the attributes `offsetOfImageCenterX`, `offsetOfImageCenterY` and `offsetOfTextX`, `offsetOfTextY`. Using one or two regions leads to the same result in the context of displaying the content, but in the former case we lose the possibility to set different actions in the case of activation or reaction.

It is also possible to combine both techniques for varying the stimulus and to create the lists of random values.

**Events**

Although GIML is already quite an advanced language, it obviously cannot cover all possible needs. Thus the mechanism of events has been implemented, in order to make it possible to connect GIML to the .NET Framework Class Library which contains a set of public methods written in any .NET framework language (e.g. C#, Visual Basic, F# or others) and to launch them in the case of any changes concerning the region states. The five possible events include: activation of a region, changing the region's state to reaction, the end of the reaction state, return to the normal state and general event launched after every change of the state of the region. Additionally, one can also associate the method with switching of the scenes.

* * *

Obviously the above description of GIML is far from complete. The tables available at the web page (Matulewski, 2019) contain most of the GIML keywords (names of elements, their attributes and predefined values) specified at the current stage of development (version 1.3). Our aim was to convince the reader that creating gaze-based applications using GIML is similar to creating web pages using HTML, and one can learn to use it quite quickly even without being a professional programmer. GCAF, the interpreter of GIML, can be downloaded from
*http://www.fizyka.umk.pl/~jacek/gamelab/download/*.

## 5. THE GIML INTERPRETER

*Gaze Controlled Application Framework* is the platform for running the gaze applications written using GIML. It is responsible for connecting with an eye tracker device, optional gathering of data at the client's end, sending messages to the server of the device if possible, recording the screen content, and of course displaying the dynamic content of GIML application. GCAF is implemented using C# programming language and Visual Studio Ultimate 2013 integrated development environment. GCAF is available for computers with operating systems Windows XP, 7, 8 and 10. It requires .NET Framework 4 Client Profile or newer, which is distributed with Windows. It can also be run on computers with Linux, using Wine and Mono Framework, but most eye tracker manufacturing companies provide drivers only for Windows. At the moment GCAF supports devices manufactured by SMI, Tobii Gaming, EyeTribe and devices implementing Open Eye-Gaze Interface (it was tested using eye trackers from Mirametrix and GazePoint).

**Figure 14.** GCAF Platform Configuration Console; on the left - "Shortcuts" tab (see text for description), in the middle - the eye tracker configuration tab, and on the right - the GIML application launch tab

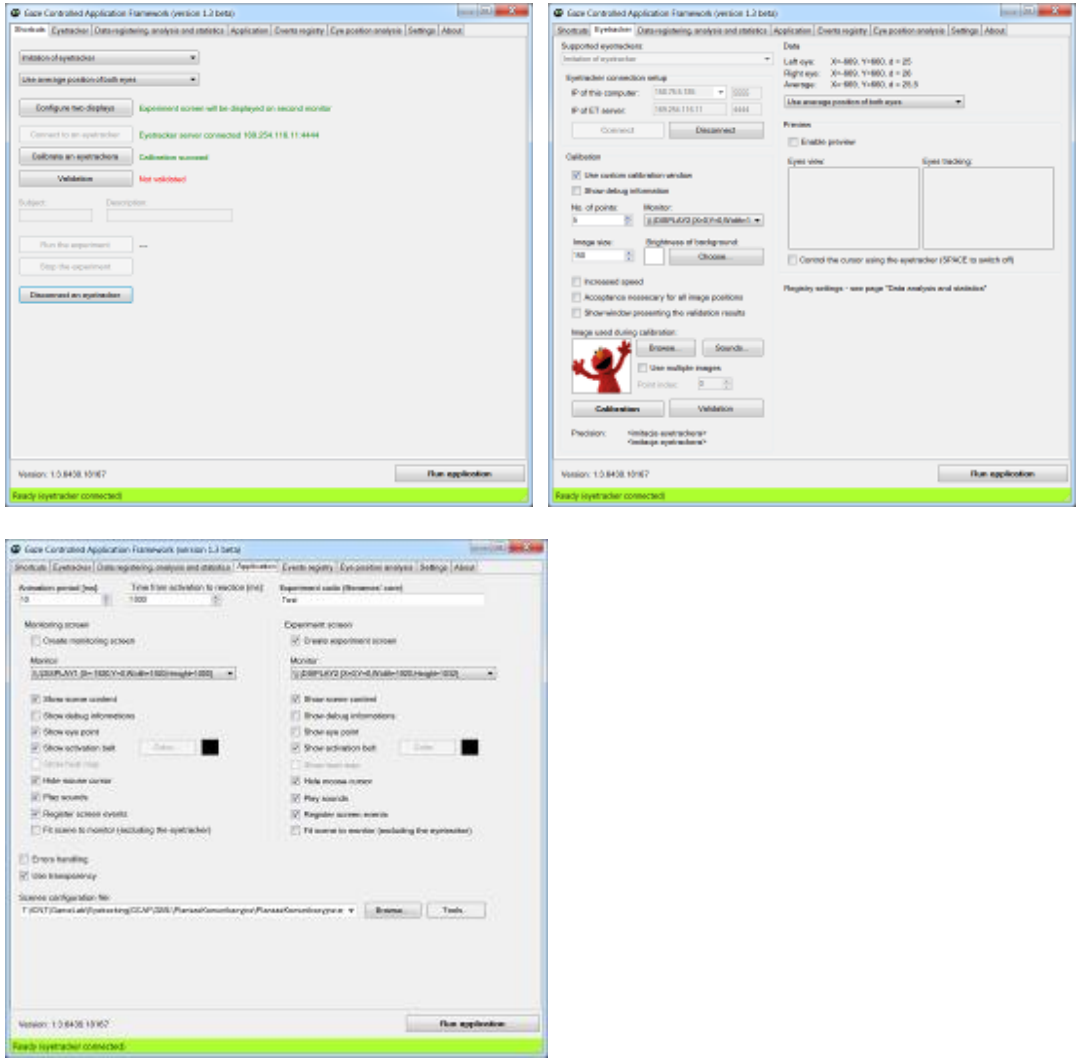

   When the GCAF platform is launched, a window visible in Fig. 14 (left) appears, allowing to select and connect the eye tracker, to calibrate and validate it, select the data to be saved while the application is running and finally to run the GIML application. In addition to collecting and recording the data, the program can analyze them "on the fly" by detecting the basic events: fixations and saccades. It is also possible to record the complete content which is shown to the user during an experiment as a movie. Once launched, the GIML application occupies the entire screen. Optionally a blank screen of any color may cover other screens to prevent the subject's attention from being attracted by their content. The GIML application can be terminated from the configuration console (if multiple screens are used) or simply by pressing the *Escape* key on the keyboard. As mentioned earlier, the GCAF can locally record the data collected while running the GIML application. These include: a sequence of screenshots or a movie, data from the samples sent by the eye tracker (gaze position, pupil diameter and others) as well as some basic statistics for regions of interest defined automatically using the GIML region positions. Logs including the sequence of events (changing the state of regions or switching the scenes) may also be saved. All files created by GCAF are text files with comma separated values (CSV format). Their paths are set in the GCAF configuration console (Fig. 14, middle). The GCAF, simultaneously to the experiment window displayed to the subject, can also show an additional control window for the researcher (cf. Fig. 14, right for its configuration). For a beginner, the easiest way to execute GIML application in the GCAF is to use the "shortcuts" tab (see Fig. 14, left). It contains a set of buttons and combo boxes that form a kind of check list before running the application. Some unique features of GCAF are also worth mentioning. For example, it delivers its own calibration window (working with selected eye trackers supporting it), where animated images can be used with an additional sound attached. It is very useful in the case of experiments with infants and children. Another feature makes it possible to use the mouse as an imitation of eye tracker, which makes the gaze application designing easier and allows to examine the mouse movement, too.

## 6. EVALUATION

The Gaze Interaction Markup Language was designed as a descriptive programming language which can also be used by non-programmers. Thus the structure of its fundamental elements, sub-elements and attributes should be easily understandable, its names should be self-explanatory and its basics should be possible to learn in several hours. By the basics we understand: defining the scenes and regions, using the states of region, declaring and using the resources and the fundamentals of navigation between scenes; all that has been briefly presented in section 3.

**Usability study procedure**

In order to evaluate the usability of GIML, we have performed a series of tests using a specified coaching method (Nielsen, 1993). Unlike in classic usability tests, subjects were encouraged to ask questions, and the moderator helped them by providing appropriate suggestions. The coaching method helped us to understand the process of learning the GIML and allowed us to identify typical problems and misunderstandings, so that really helpful documentation could be designed later. In each test both the screen capture and the participant's face were recorded. The analysis was performed using the so-called rainbow spreadsheet proposed by Tomer Sharon (Sharon 2012, Sharon 2013, Viviano 2014), designed to analyze video recording of subject-coach interactions. Twenty-one individuals took part in the study, 11 of them study Informatics at the Faculty of Physics, Astronomy and Informatics of Nicolaus Copernicus University (Group I), while 12 study Cognitive Science at the Faculty of Humanities of the same university (Group K). In both groups, the proportion of men and women roughly corresponded to sex ratio of all students at these faculties (see Fig. 15). The participants were between 20 and 26 years of age. All participants reported normal vision, in some cases subject to corrective devices. Results of two individuals from Group K were excluded from the analysis. One student was extremely excited and nervous from the beginning of the session; in the second case symptoms of physical and mental fatigue were noticed (frequent eye blinking, rubbing the face, complaints about contact lenses etc.). Ultimately, results of only 10 persons from Group K were taken into account.

**Figure 15. Groups of participants**

| Group | All | I | K |
|---|---|---|---|
| Number of people | 21 (23) | 11 | 10 (12) |
| Number of women | 9 (10) | 2 | 7 (8) |
| Number of men | 12 (13) | 9 | 3 (4) |
| Percentage of women in groups | 57% | 8% | 70% |
| Percentage of women at the faculty | | 10% | 73% |

The survey was performed for the Polish version of GIML. The mouse-based eye tracker imitation was used in order to achieve similar conditions for all the subjects, without the inconvenient procedure of calibration and a need for recalibrations. We believed the use of eye tracker imitation would not be of any consequence since the aim of the study was to examine the process of learning the GIML language, rather than the use of GIML in interactive applications. Recording of the desktop content and the view of the subject's laptop built-in camera as well as the Camtasia Studio 8 were used in the procedure. The sessions were performed by the subjects sitting at separate desks in silence, using Lenovo Y550P laptop working with Windows 10 operating system and the USB microphone Novox NC-1.

The sessions consisted of several stages: 1) an introduction, during which the investigator explained the aim of the study i.e. the evaluation of GIML, 2) a learning phase during which the subjects read a printed text presenting the fundamentals of GIML, 3) the questionnaire gathering information on the subjects' programming experience, 4) a hands-on tutorial on GIML that explained how to define the regions, understand their states, and scene navigation, 5) tasks examining the knowledge of GIML, and 6) the questionnaire on the subjects' personal opinions about the GIML. During the examination (stage 5 from the above list) the subject received a sheet of paper containing a list of tags and attribute names and some of their predefined values necessary to perform the tasks. At that time he/she was not allowed to look at the tutorial or the code files written earlier. The test (Stage 5) started with two tasks checking understanding of the GIML code. In the first one the code presented to the subject contained a single scene definition and a single region with all three states defined. Each state had different text and color (Fig. 16). The aim of this task was to check whether the subject understands the idea of region states. The second task checked the understanding of navigation between scenes and understanding of using the resources. This time the code presented to the subject (Fig. 17) contained two scenes. In the first one the region similar to that in the first task was used, but texts were replaced by images. Thus the resource declaration elements were also added. The transition of the region into a reaction state caused switching to the second scene containing two regions: one with the image and one with the label "the end". In both tasks the subject was asked to describe what would be visible on the screen, also if the gaze position entered the regions and stayed there longer. In order to judge the knowledge of GIML, we constructed a simple measure taking into account selected facts, which should be noticed and expressed by the subjects. All of these facts had weights attributed to them (negative in the case of mistakes). Examples of such facts for the first task are shown in Fig. 18.

**Figure 16. Code presented to the subject during the first task (English translation)**

```
<?xml version="1.0" encoding="utf-8"?>
<settings language="en">
  <scenes nameOfDefaultScene="first"
        originalScreenSizeX="1024" originalScreenSizeY="768">
    <scene name="first" backgroundColor="Beige">
      <region name="region1" shape="rectangle"
            locationOfCenterX="150" locationOfCenterY="100"
            sizeX="200" sizeY="100"
            text="Navy"
            font="Times" fontSize="30" fontColor="Navy">
        <activation text="Blue"
                font="Times" fontSize="30" fontColor="Blue" />
        <reaction text="Cyan"
                font="Times" fontSize="30" fontColor="Cyan" />
      </region>
    </scene>
  </scenes>
</settings>
```

**Figure 17. Code presented to the subject during the second task**

```
<?xml version="1.0" encoding="utf-8"?>
<settings folder="C:\GIML\UX_Study\Assets" language="en">
  <images>
    <image name="green_disk" path="imgs\green.png" />
    <image name="yellow_disk" path="imgs\yellow.png" />
    <image name="red_disk" path="imgs\red.png" />
  </images>
  <scenes nameOfDefaultScene="default"
        originalScreenSizeX="1024" originalScreenSizeY="768">
    <scene name="default" backgroundColor="white">
      <region name="disk" shape="rectangle"
            locationOfCenterX="150" locationOfCenterY="150"
            sizeX="200" sizeY="200"
            nameOfImage="green_disk">
        <activation nameOfImage="yellow_disk" />
        <reaction actionType="transitionToScene" nameOfTargetScene="end" />
      </region>
    </scene>
    <scene name="end" backgroundColor="Black">
      <region name="disk" shape="rectangle"
            locationOfCenterX="150" locationOfCenterY="150"
            sizeX="200" sizeY="200"
            nameOfImage="red_disk" />
      <region name="caption" shape="rectangle"
            locationOfCenterX="180" locationOfCenterY="300"
            sizeX="200" sizeY="50"
```

```
                    text="end" fontColor="White" fontSize="30" />
        </scene>
      </scenes>
    </settings>
```

**Figure 18. Facts monitored and evaluated during the first task**

| Facts | Weight | All | I | K |
|---|---|---|---|---|
| "the color of the background is beige" | 1 | 38% | 36% | 40% |
| "one scene and one region" | 1 | 43% | 64% | 20% |
| "region displays text" | 1 | 95% | 100% | 90% |
| the content of the text (also if asked) | 1 | 86% | 82% | 90% |
| all font properties were listed | 1 | 33% | 36% | 30% |
| changing the text and its color in the case of changing the state of the region | 1 | 100.00% | 100% | 100% |
| giving the precise colors of text after changing the state of the region | 1 | 90% | 100% | 80% |
| the appearance of the region in the normal state was skipped and only activation and reaction states were described | -1 | 29% | 9% | 50% |
| **Points (scale from -1 to 7)** | | **4.57** | **5.09** | **4.00** |
| **Rescaled points (0 to 3)** | | **2.09** | **2.28** | **1.88** |

In the third task the subject was shown the working application with one region displaying the text depending on its states. He or she was asked to write the GIML code starting from an empty file, reproducing the appearance and behavior of the presented application. We measured how often the subject looked at the printed list with the GIML keywords, what questions he/she asked and how many times. We also counted mistakes, requests for help, researcher's answers and the number of activities providing assistance in code writing. As previously, weights were attributed to all observations and a numeric value was calculated to reflect the subject's ability to write a correct GIML code.

The final task required each subject to write a gaze-controlled photo-viewer. The subject was only provided with a description of the application behavior (with a screenshot visible in Fig. 19) which was introduced by the investigator. All the necessary resources (three photos and two images of arrows) were also provided. The screenshot contains three regions: the large one in the middle for the photo and two arrows on the left and the right side of the photo. A gaze fixed at an arrow for more than one second should cause a change of the photo. Like in the case of the third task, all questions, comments and requests for assistance were recorded. The subject was still allowed to use a printed list of GIML keywords.

**Figure 19. Screenshot of application which a subject was supposed to write in the fourth task**

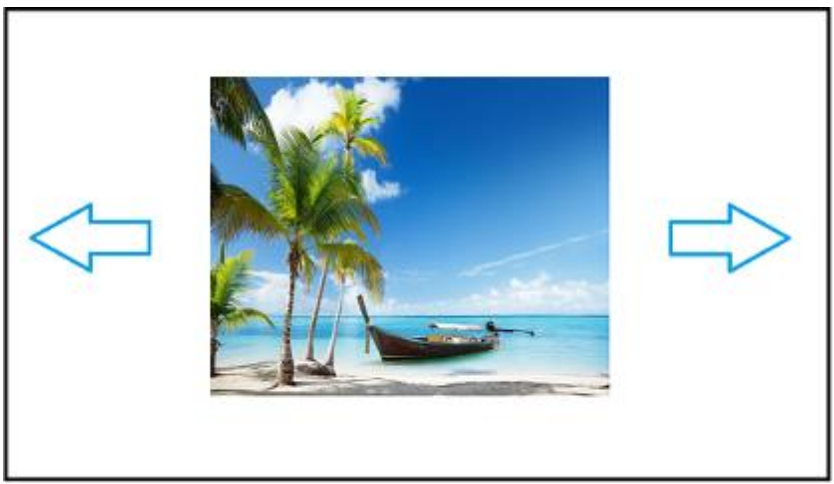

**Results and discussion**

All the participants from both groups completed all the four tasks designed to test their knowledge of GIML. While only two students from Group K required more extensive assistance, the others succeeded without such help. This is the most important observation of the presented study. It shows that GIML can be learned not only by professional programmers but also by almost anyone. It certainly takes longer for non-programmers to learn their first programming language, even a descriptive one, than for IT students to learn the next one, but in the case of GIML the former was still possible in a relatively short time. The non-programmers also made more mistakes, which was expected since in general they are not used to writing source codes with their precise requirements. Nevertheless they were able to learn GIML and to write a proper GIML code.

The participants were divided into two groups based on their course of study. These two groups differed significantly in terms of their knowledge of HTML, declared ($U_{M-W}$ = 2, p < .001) and actual ($U_{M-W}$ = 20, p < .023). For example, we asked about the meaning of the A element in HTML. 60% of the subjects from Group I answered correctly, while 100% subject from Group K did not know the correct answer ($\chi^2$ = 8.57, p = .003). In addition, some experience with web programming was declared by 90% of

Group I compared to only 10% of Group K ($\chi^2$ = 12.8, p < .001). Thus, it is evident that the division of the subjects into two groups reflects the differences in their programming skills.

The points, rescaled to 0-3 range, scored in all the tasks at Stage 5, together with the total points and total time during which the tasks were completed, are presented in Fig. 20. The number of points measures the proficiency in using GIML and is indicative of the number of errors made by the subjects (cf. Fig. 18). One can notice the time needed for performing the tasks was approximately twice as long in the case of the Cognitive Science students compared to the IT students. Since there was no time pressure during the session, this should more or less reflect the time differences in learning GIML in "home conditions" by individuals with and without technical background. Moreover, with an increasing time needed for completing tasks 3 and 4, where the subjects had to write the GIML code, the number of points gained decreases for all the subjects (correlation in task 3: $\rho$ = -.71, p < 0.01 and in task 4: $\rho$ = -.72, p < 0.01). The clear difference between both groups in the total execution time suggests that it is worth checking this correlation in both groups. For Group I such correlation was not observed, while for Group K it was strong. However, it should be pointed out that the size of both groups, considered separately, was insufficient to obtain reliable statistics. Analysis of the video recorded during the study shows that a longer task execution time in Group K was mostly related to search for errors in the code, mainly typos, typical for individuals learning to program and it was only rarely associated with thinking about the code to be written. The duration of the task performance also correlates with the declared level of GIML difficulty ($\tau$-C = .53, p < .001). Also in this case the correlation was significant in Group K ($\tau$-C = 0.72, p = .001), while it was absent in Group I ($\tau$-C = 0.3, p = .396).

**Figure 20. Results of tests in rescaled points**

| Facts | All | I | K |
|---|---|---|---|
| Task 1 ("reading the GIML code – regions and states") | 2.09 | 2.28 | 1.88 |
| Task 2 ("reading the GIML code – resources and navigation") | 2.21 | 2.30 | 2.11 |
| Task 3 ("reproducing the application – regions and states") | 2.01 | 2.39 | 1.60 |
| Task 4 ("the photo viewer") | 1.66 | 2.42 | 0.83 |
| Total points (scale from 0 to 12) | 7.97 | 9.39 | 6.41 |
| Overall time for performing tasks | 48:00 | 37:30 | 59:33 |

We have also analyzed the results of the questionnaire filled in by the subjects at the end of the examination session (Figs. 21 and 22). Subjects from Group K declared that GIML is more difficult than the subjects from Group I ($U_{M-W} = 18$, $p < .028$). Generally all the subjects reported that GIML is easier than HTML (55% positive versus 15% negative, while 30% had no opinion) and that using a single file for multiple scenes (contrary to HTML) is helpful (75%). Interestingly, the Polish version of GIML keywords (the study was done for Polish language native speakers) was helpful for non-programmers (70%) and at the same time it was rather disturbing for some programmers (only 39% subjects from Group I claimed it was helpful). The keywords written in Polish were evaluated as inconvenient by 27% of the programmers in contrast to 12% of the non-programmers. All the subjects declared that keyword names correctly suggested their meaning (100%).

**Figure 21. Results of the final questionnaire (positives)**

| Questions | All | I | K |
|---|---|---|---|
| What is the difficulty of GIML? (scale 0-easy, 1-medium, 2-difficult) | 1.0 | 0.6 | 1.4 |
| Is GIML easier than HTML? (only if the subject declared a knowledge of HTML) | 79% | 80% | 75% |
| Are Polish keywords helpful? | 55% | 39% | 70% |
| Do keyword names convey their meaning? | 100% | 100% | 100% |
| Is it helpful to use a single file for many scenes? | 87.5% | 80% | 95% |

**Figure 22. Results of the final questionnaire (negatives)**

| What was inconvenient? | All | I | K |
|---|---|---|---|
| typos | 29% | 45% | 14% |
| using Polish language (Polish keywords) | 19% | 27% | 12% |
| setting region location using center point | 5% | 9% | 1% |
| XML syntax, need for putting end tags | 29% | 18% | 37% |
| quotation marks | 5% | 0% | 9% |
| assigning attributes to proper tags | 10% | 9% | 10% |

Based on the analysis of the videos recorded during the study, we also concluded that some code editor with GIML keyword suggestions would be helpful. This should reduce the need for checking the printed list of GIML keywords that was noticed in both groups. In this editor the XML header should also be automatically added at the start - it caused many problems to some subjects. A continuous preview of the edited scene would also be helpful, since it was observed the subjects from Group K wrote the code for longer time without trying to run the application. This leads to accumulation of typos and other errors, the search of which greatly increases the total time needed for writing the code. It should be mentioned that after a pilot-study, which was done for two subjects with no programming knowledge, we introduced case insensitivity into GIML (in contrast to XML).

## 7. EXAMPLES OF USE

Two different examples of using GIML applications will be briefly presented below. The first one relates to communication with infants, for whom gaze, besides crying, is the only natural way of "sending messages" to their caregivers. The second one relates to the so-called communication boards for disabled people, for whom gaze is also the only means of communication.

**Cognitive experiments on language acquisition in infants**

Communication requires interactions of those involved. Infants, even at an early stage of their development, are not just passive recipients of the messages, but also senders, of course at the beginning only of non-verbal messages which usually are clearly legible for their parents (eg. Leclère et al., 2014; Ramírez-Esparza, García-Sierra, & Kuhl, 2014; Hamilton, Southgate, & Hill, 2016). Research on language acquisition and perception of speech at such an early stage of life is fundamental for our understanding of human speech development and for identifying the most important learning process factors in general. Infants in the preverbal period are capable of differentiating speech sounds from both their mother tongue and a foreign language (Kuhl, 2003). By the age of eight months babies are able to distinguish all sounds from any language, but then this ability gradually disappears. It is possible, however, to temporarily maintain sensitivity to foreign phonemes, which in the future should result in easier language learning (Kuhl et al., 2003, 2004, 2011, 2015). Experiments based on this assumption are currently being carried out in our laboratory, with the use of training and test procedures designed with GIML. The procedures used in these experiments are based on the ability to predict the position of an object appearing on the screen (known as anticipatory movement paradigm) contingent on the previously given sound stimulus (see Albareda-Castellot et al, 2011; Bjerva et al., 2011). For example, in the first diagnostic method applied in our laboratory, in which GIML application was used, we follow the experimental procedure proposed by Albareda-Castellot et al. (2011). The child sees a drawing of an Elmo character at the bottom of the screen. The size of the image changes periodically drawing the infant's attention. Eyes directed at the drawing start pseudorandom playback, while the drawing hides behind a T-shaped cover (Fig. 23). After four seconds the drawing appears at the left or at the right arm of the covering region. The side depends on the speech sound presented earlier. Measurement of the baby's gaze direction (see the areas of interest, AOIs shown in Fig. 23) is performed before the figure of Elmo appears from behind the cover, thus revealing the side expected by the child (anticipation). That in turn allows to determine how well babies can discriminate similar phonemes. The entire experiment consists of 24 scenes, including 18 scenes with a sound stimulus and four scenes attracting the baby's attention with an animated object in the center of the screen that disappears when the baby's eyes focus on it for 300 ms.

**Figure 23. The procedure of Albareda-Castellot et al. (2011). The gaze aware region initiating the trial and areas of interest are marked with the dashed lines**

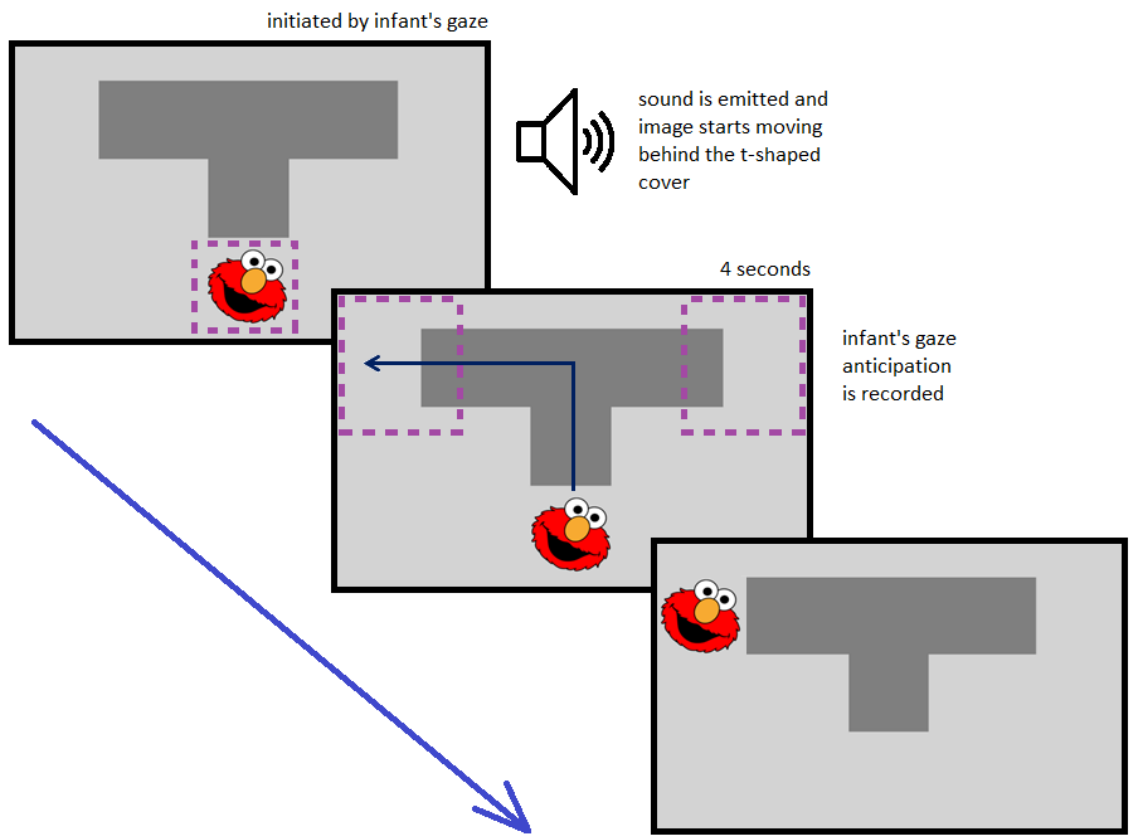

A number of other experiments have also been conducted in which the paradigm presented above was modified. For example the experiment originally proposed by McMurray and Aslin (2004), and designed for six-month infants, comprises a series of four trials based on two alternative forced choice paradigms (2AFC). A similar experiment was also implemented using GIML, following Bjerva et al. (2011). In this case there are two areas where the drawings are shown on the screen (Fig. 24). At the first stage of trial both areas are empty and an animated picture appears between them in order to draw infant's attention to the screen. Then the picture disappears and the baby hears a sound stimulus, and after 1200 ms the figure appears in one of the areas. Like in the first method, the direction of the gaze is measured before the picture appears. In this experiment 36 scenes are displayed, 18 of which are in line with the description above, and all the others are intended to attract the child's attention. In addition, every fourth soundstage scene is lacking the visual stimulus. Observation of the child while he/she is waiting for the visual object and looking at the appropriate area makes it possible to obtain the best possible confirmation of speech sounds differentiation.

**Figure 24. The screenshot from the procedure of the two-choice experiment**

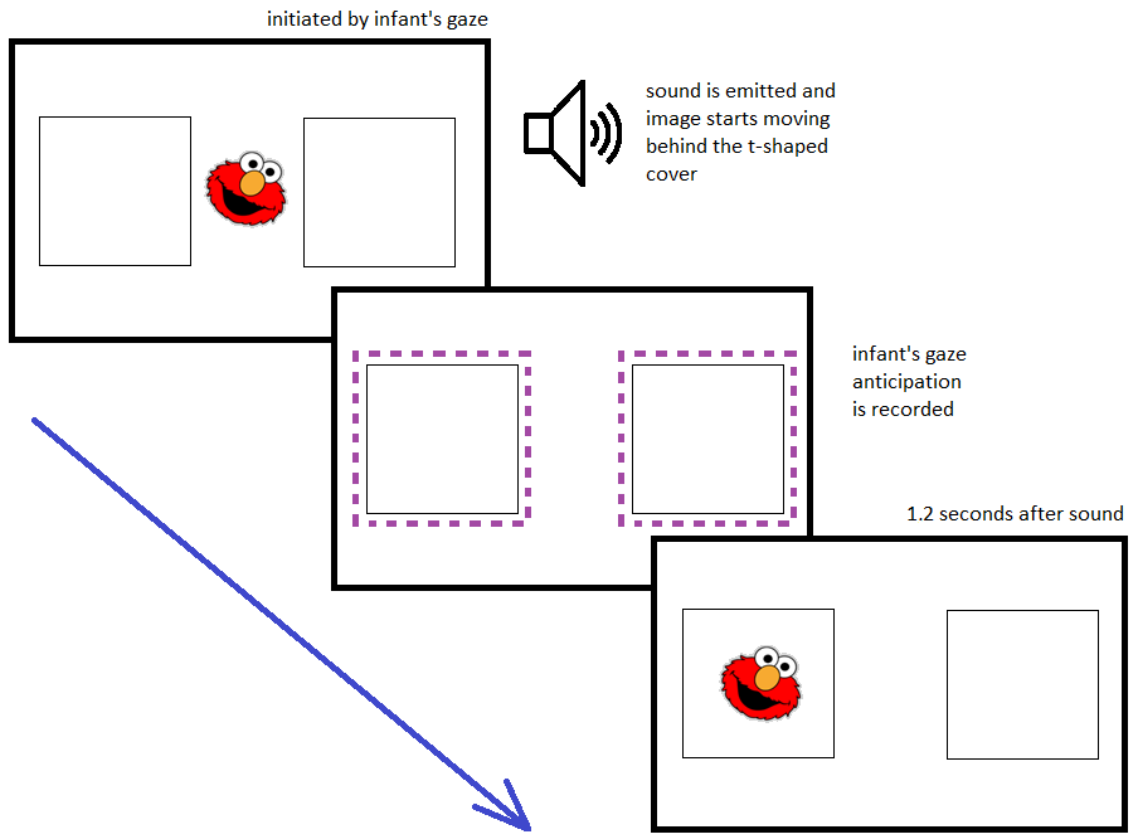

The training part of the study also includes two applications prepared using GIML and run on the GCAF platform. In the first we used the scheme proposed by Wang et al. (2012). The original experiment comprised a presentation of a series of photographs of animals that slowly evolved changing one to the other. However, this process becomes much faster if infants focus their eyes at the red dot visible at the right side of the picture. Hence, the dot works as a kind of a switch. We modified this experiment to design the training stage, by replacing images of animals with fragments of cartoons with French dubbing (Dreszer, et. al., 2015). Thanks to the dot, the child had the choice: to watch the currently playing part of the movie or to switch to the next one (Fig. 25).

**Figure 25. In this training infant can scroll movies by staring at the right red dot**

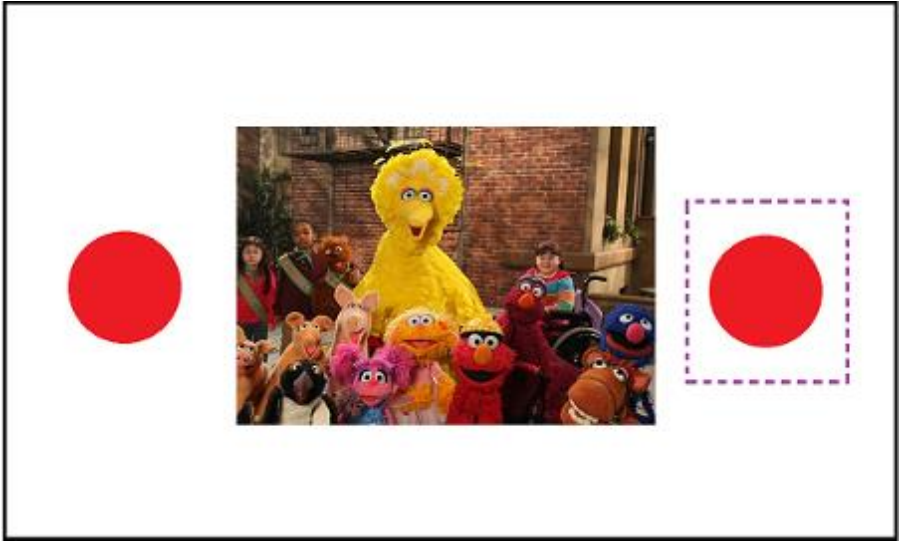

Our original experiment is based on an interactive fairy tale that is controlled by a child's gaze (Fig. 26) (Szmytke, et. al, 2018). It uses pictures and sounds (in French) from the *Peppa Pig* movies. There are three elements in each scene. The infant activates them by looking at one of these elements, causing rotation of the round colorful borders. When the infant is looking for a longer time (state of reaction) at one character, the entire scene outside the selected element is covered by a translucent veil and the character in this area begins to tell its story (spoken in the infant directed speech). In this training the child chooses the character he/she wants to listen to, but then the characters and their elements disappear. Based on the literature (cf. Kuhl et al., 2011, Tomalski et al., 2013, Hamilton, Southgate & Hill 2016), we expect that the ability to interact with a fairy tale will result in better training results than passive training.

**Figure 26. Interactive cartoon. On the left, two of the three interactive elements are visible, on the right – one element is activated, the sound can be heard and the blackout covers the rest of the scene.**

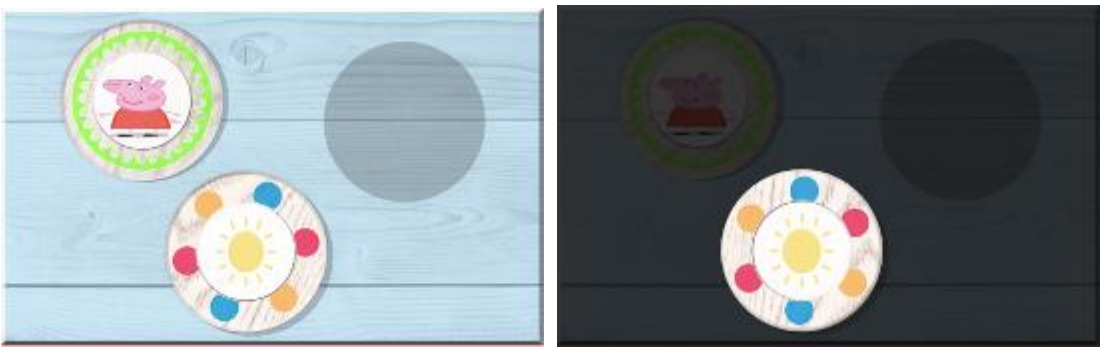

**Communication boards**

Another area of GIML usage is augmentative and alternative communication (Bates et al., 2007; Hill 2010; Townend et al., 2016). Using GIML we digitalized the gaze communication boards designed by Zawadzka and Murawski (2015) for everyday communication with patients who do not control their body muscles, but control their eyes. Every message involves a two-step procedure: firstly, the category needs to be selected, and secondly, a particular message is chosen. Dwell-time selection method is used in both steps. All boards have a fixed scene structure, where four images with captions are presented at the middle of every screen edge. Additionally two bottom corners may contain regions allowing to switch between the boards at the same level (scrolling through categories or messages), the region in the upper left corner moves back and the region at the upper right corner cancels the selected message (see Fig. 27 for samples). The centers of all the scenes are insensitive to gaze and allow the user to rest. Eleven categories, including "I am ...", "I want ..", "Pain", "Emotions", "Free time" and others, are represented by images in the three scenes. Every category has a color attributed and the same color is used as a background in the corresponding scenes with messages from this category, providing a clear feedback. After the category is selected, one can scroll through the scenes with particular messages and select one. Such selection causes all the other regions to disappear and the specific voice message (recorded by a speaker) is played. One can choose between a male or female voice by changing the resources folder in GIML code. The choice of region can be canceled or the communication activity may be finished by returning to the category scenes. A frequently used board with common answers ("Yes", "No", "I don't know", "I don't understand") is also available, as well as a scene with dark background and schematic image of eyes closed, which can be selected by the patient to signal the need for a break. The system also includes a simple test for using the boards. Three scenes with colors, common objects and animals are presented to the user. After entering the scene, the name of one of the colors, objects or animals is spoken and the user is asked to select the correct region.

**Figure 27. Digitalized communication boards (in Polish) prepared using GIML. Two columns contain two paths of choices to be made by the user. The first row contains sample category sheets, the next carries specific messages and the last – confirmed message**

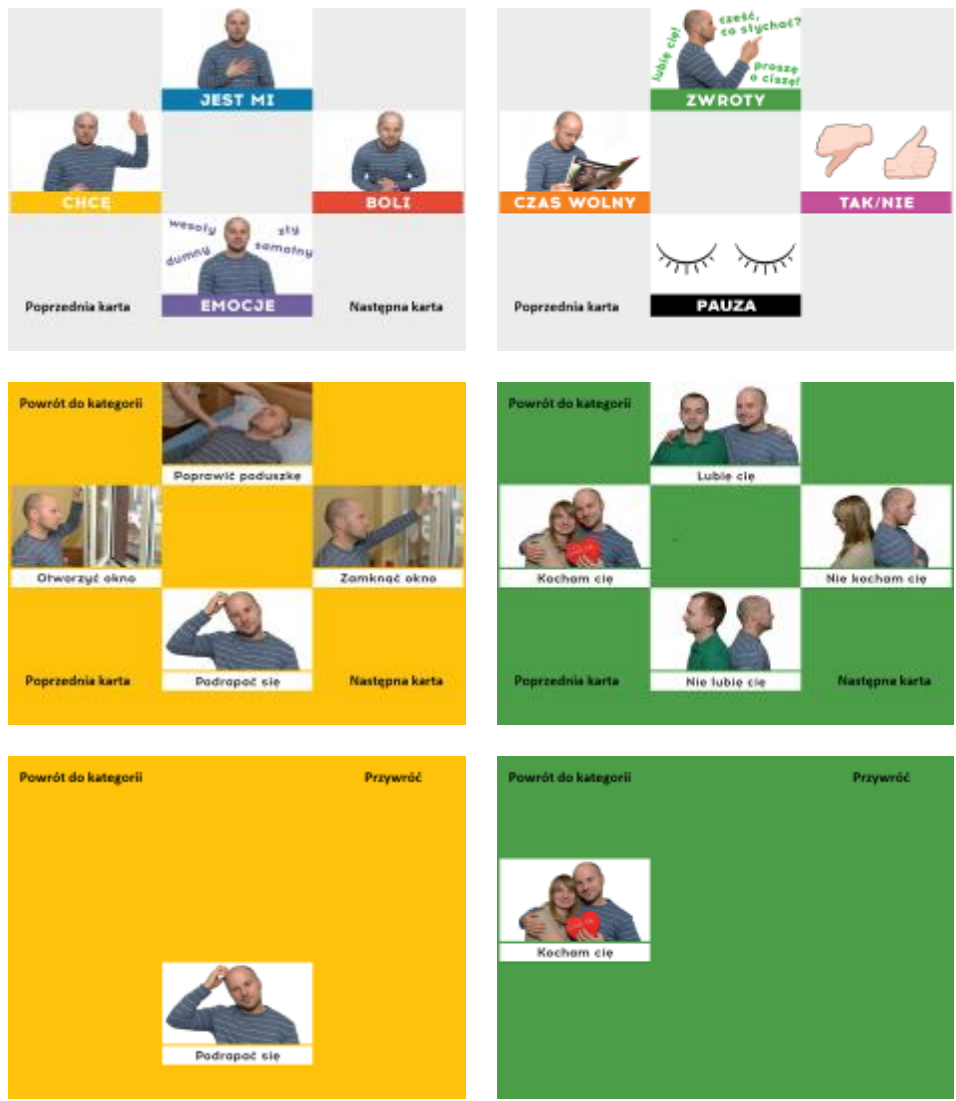

Because the design of the boards and all the photos were already prepared in printed version there were no additional costs of digitalization and it took only a few hours to create the ready-to-use GIML program. Notably, further modifications, extensions or personalization e.g. with self-made photos of user or their family members, additional messages matching the user's special needs or categories related to their hobby, may be done easily using the same scene schema. Comparison of user experience related to printed and digitalized boards is now being carried out. We are also interested in the possibility of adjusting the boards to preferences based on temperamental and personality traits.

We have also thought about other applications for GIML and GCAF. For example we have explored the possibility of creating games using gaze interaction as a control method. A simple example is the so-called hidden objects game which even in its original form demands strong gaze engagement. We use dwell-time for pointing to the regions that users focus on. After such regions are identified, they are removed both from the scene and from the list at the side. In addition the game can be made more difficult by using the so-called flashlight effect, which cuts out the peripheral vision. We have also developed a kind of first-person perspective (FPP) game named "Zombie". In this game the scene backgrounds imitate a perspective view of corridors. Along these corridors zombies are approaching the user (their regions are enlarging). The user's avatar cannot move, but fortunately has the ability to burn zombies by directing its gaze. Since the zombies are continuously approaching the time pressure is important. At every level several zombies must be defeated, which is followed with the arrival of the "boss", who is faster and more resistant (Szwarc, 2017).

## 8. SUMMARY AND FUTURE PLANS

The usability study shows that it is possible to learn the basics of GIML in about an hour. Of course we are aware that learning more advanced techniques requires much more time, but it is still achievable for an average user, especially with the help of someone with technical background. That creates the possibility for creating personalized applications for disabled people which can improve the quality of their lives without extensive costs for program development. Delivery of such a tool was our main goal. As mentioned above, we currently work on several new features. One of these is a special region for presenting the text, which automatically defines the areas of interest for letters, words, sentences and paragraphs. Later, we would like to provide GIML with the option for controlling the application by using blinks as an alternative to dwell-time method for accepting the choices (the use of keyboard buttons is already possible). Furthermore, we want to supplement the element `region` with new sub-elements describing the situation when the left or right eye is closed for a given time interval while the second eye points to the region. We want to provide the `scene` element with attributes activated if both eyes are closed. We also work on merging the GCAF and a text entry system that we have developed, operating with several kinds of gaze entry systems. We also want to add offset correction (one-point recalibration procedure allowing to reduce an accuracy error) as a GIML feature. Currently this software is available only in English and Polish language. More supported languages in the GIML, as well as in the GCAF configuration window will be added for users from different countries.

# LISTS OF GIML ELEMENTS AND ITS ATTRIBUTES

Jacek Matulewski, e-mail: *jacek@phys.uni.torun.pl*
Version 1.3 (2019-11-20)
Most recent version available at: *http://fizyka.umk.pl/~jacek/gamelab/download/gcaf/GIML_Keywords.pdf*

    Below the most important GIML elements with its all attributes are listed in four supported languages. Attributes marked with bold font are obligatory (see also column "Required"). Last column contains sample values. They are listed by comma or in separate lines. In case of optional attributes, the default values are marked with bold font.

Tab. A.1. Attributes of `setting` element

| settings / paramètres / einstellungen / ustawienia ||||||||
|---|---|---|---|---|---|---|---|
| English | French | Deutsch | Polish | Req. | Interpretation | Default (bold) and sample values |
| `folder` | `dossier` | `verzeichnis` | `katalog` | no | directory of assets file on hard disk | C:\Users\Jacek\Assets |
| `language` / `sprache` / `langue` / `język` |||| no | language code | **en**, fr, de, pl |
| `library` | `bibliothèque` | `bibliothek` | `biblioteka` | no | path to .NET library containing event methods | C:\Users\Jacek\Library.dll |

Tab. A.2. Attributes of `image` element

| image / image / bild / rysunek | | | | | | |
|---|---|---|---|---|---|---|
| English | French | Deutsch | Polish | Req. | Interpretation | Sample values (default is bold) |
| **name** | **nom** | **name** | **nazwa** | yes | identifier | img1 |
| **path** | **chemin** | **pfad** | **ścieżka** | yes | path of asset file | images\img1.png images\img1\*.bmp |
| transparencyKey | cléTransparence | transparenzSchlüssel | kluczPrzezroczystości | no | color which will be changed to transparent | White **Fuchsia** #FF00FF #00FF00FF |
| runningPeriod | duréeFonctionnement | abspiellaufzeit | okresOdtwarzania | no | time (in ms) of showing full animation (in case of GIF images or sets of images) | **1000** |
| runFromFrame | lancementDepuisTrame | vonFrameAbspielen | odtwarzanieOdRamki | no | starting frame (in case of animation) | **0** |
| runToFrame | lancementJusquaTrame | zuFrameAbspielen | odtwarzanieDoRamki | no | ending frame (in case of animation) | 10, 100, **-1** (*run to end*) |

| `keepInMemory` | `garderEnMémoire` | `cacheSpeicher` | `przechowujWPamięci` | no | loads whole series of images into memory (in case of animation) | en: yes \| **no**<br>fr: oui \| **non**<br>de: ja \| **neine**<br>pl: tak \| **nie** |

Tab. A.3. Attributes of `sound` element

| sound / son / ton / dźwięk | | | | | | |
|---|---|---|---|---|---|---|
| English | French | Deutsch | Polish | Req. | Interpretation | Sample values (default is bold) |
| **name** | **nom** | **name** | **nazwa** | yes | identifier | snd01 |
| **path** | **chemin** | **pfad** | **ścieżka** | yes | path of asset file | sounds\snd01.wav |
| repetitionNumber | numéroRépétition | wiederholungszahl | liczbaPowtórzeń | no | number of times the whole sound sample is played | **1**, 3, 10 |
| inBackground | enArrièrePlan | imHintergrund | wTle | no | sound is played to the end even if next sample is played | en: yes \| **no** <br> fr: oui \| **non** <br> de: ja \| **neine** <br> pl: tak \| **nie** |
| volume | volume | lautstärke | głośność | no | sound playing volume | 0.1, 0.5, **1**, 2 |

Tab. A.4. Attributes of `movie` element

| movie / film / film / film | | | | | | |
|---|---|---|---|---|---|---|
| English | French | Deutsch | Polish | Req. | Interpretation | Sample values (default is bold) |
| **name** | **nom** | **name** | **nazwa** | yes | identifier | mov01 |
| **path** | **chemin** | **pfad** | **ścieżka** | yes | path of asset file | movies\mov01.avi |
| transparencyKey | cléTransparence | transparenzSchlüssel | kluczPrzezroczystości | no | color which will be changed to transparent | White **Fuchsia** #FF00FF #00FF00FF |
| repetitionNumber | numéroRépétition | wiederholungszahl | liczbaPowtórzeń | no | number of times the whole movie is played | **1**, 3, 10 |
| inBackground | enArrièrePlan | imHintergrund | wTle | no | sound is played to the end even if next sample is played | en: yes \| **no** fr: oui \| **non** de: ja \| **neine** pl: tak \| **nie** |
| volume | volume | lautstärke | głośność | no | sound playing volume | 0.1, 0.5, **1**, 2 |

Tab. A.5. Attributes of `scene` element

| scene / scène / szene / scena | | | | | | |
|---|---|---|---|---|---|---|
| English | French | Deutsch | Polish | Req. | Interpretation | Default (bold) and sample values |
| **`name`** | **`nom`** | **`name`** | **`nazwa`** | yes | identifier | scene01 |
| `backgroundColor` | `couleurArrièrePlan` | `hintergrundfarbe` | `kolorTła` | no | color of scene background | **White** Fuchsia #FF00FF #00FF00FF |
| `nameOfBackgroundImage` | `nomDImageArrièrePlan` | `nameDerHintergrundZeichnung` | `nazwaRysunkuTła` | no | identifier of image displayed in background of scene | img01 |
| `nameOfBackgroundSound` | `nomDeSonArrièrePlan` | `nameDesHintergrundklangs` | `nazwaDźwiękuTła` | no | identifier of sound sample played after enter to scene | |
| `blackoutDegree` | `degréBlackout` | `blackoutGrad` | `stopieńZaczernienia` | no | transparency of blackout cover | **0**, 128, 255 |
| `blackoutColor` | `couleurBlackout` | `blackoutFarbe` | `kolorZaczernienia` | no | color of blackout cover | White **Black** #FF00FF #00FF00FF |
| `blockingRegionsDuringBlackout` | `blocRégionsDurantBlackout` | `regionblockierenBeiBlackoutt` | `blokowanieObszarówPrzyZaczernieniu` | no | switch regions to be not responsive after the blackout is activated | en: yes \| **no** fr: oui \| **non** de: ja \| **neine** pl: tak \| **nie** |
| listOfRegionsToDisable | listeDesRégionsADésactiver | regionenlisteZumAusschalten | listaObszarówDoWyłączenia | no | separated with semicolon list of regions' identifiers, which will be disabled after enter to scene | rgn01;rgn01 |
| nameOfRegionEnabledAfterAllRegionsAreDisabled | nomRégionsActivéesAprèsToutesRégionsDésactivées | nameDesAktiviertesBereichsNachDeaktivierungDerAlleAnderenBereiche | nazwaObszaruWłączanegoPoWyłączeniuWszystkichObszarów | no | identifier of region, which will be enabled, after none enabled region remains on the scene | rgn03 |
| resetAfterEnter | réinitialiserAprèsEntrée | nachEingabeZurücksetzen | resetujPoWejściu | no | after return to scene, it forget its previous state | en: yes \| **no**<br>fr: oui \| **non**<br>de: ja \| **neine**<br>pl: tak \| **nie** |
| spotlight | projecteur | taschenlampe | latarka | no | all scene are covered with black except the circle around the gaze position | en: yes \| **no**<br>fr: oui \| **non**<br>de: ja \| **neine**<br>pl: tak \| **nie** |
| spotlightRadius | rayonProjecteur | radiusDerTaschenlampe | promieńLatarki | no | radius of circle around the gaze position in which the scene is visible | 50, 100, **150**, 200 |
| nameOfListsSwitchedOverAfterEnter | nomDesListesBasculéesAprèsEntrée | namenDerListenAmEingabeGewechselt | nazwyListPrzełączanychPrzyWejściu | no | lists' identifieres which values will be switched after enter to scene | list01;list02 |
| nameOfRegionEnabledAfterListFinished | nomDesRégionsActivéesAprèsFinListe | nameDesBereichsEingeschaltetNachEndeDerListe | nazwaObszaruWłączanegoPoZakończeniuListy | no | region's identifier, which will be enabled in case any list's elements will be | region04 |

|  |  |  |  |  | depleted |  |
| --- | --- | --- | --- | --- | --- | --- |
| `onSceneChanged` | `surScèneChangé` | `nachSzenenwechsel` | `poZmianieSceny` | no | name of method from .NET library, which will be executed after enter to scene | methodScene01 |

Tab. A.6. Attributes describing state of region which can be used in `region` element as well as in its sub-elements `activation` and `reaction`

| region / région / region / obszar<br>activation / activation / aktivierung / aktywacja<br>reaction / réaction / reaktion / reakcja | | | | | | |
|---|---|---|---|---|---|---|
| English | French | Deutsch | Polish | Req. | Interpretation | Default (bold) and sample values |
| `actionType` | `typeAction` | `aktionstyp` | `typAkcji` | no | one of six action performed by region after region's proper state is entered:<br>1) no action (none);<br>2) add the border around the region (border), the color and thickness of the border are determined by attributes `borderWidth` and `borderColor`;<br>3) switch to other scene, which identifier is pointed in attribute `nameOfTargetScene` (transitionToScene);<br>4) translation to new position along the path described in `path` attribute with speed given in `speed` attribute (move),<br>5) reset the current region or<br>6) whole current scene | en: **none** \| border \| transitionToScene \| move \| resetRegion \| resetScene<br>fr: **aucun** \| bordure \| transitionVersScene \| bouger \| réinitialiserRégion \| réinitialiserScène<br>de: **keine** \| rahmen \| übergangZuScene \| abstand \| regionZurücksetzen \| sceneZurücksetzen<br>pl: **brak** \| ramka \| przejścieDoSceny \| przesunięcie \| resetujObszar \| resetujScenę |
| `borderWidth` | `largeurBordure` | `rahmenbreite` | `grubośćRamki` | no | thickness of border around | 1, 2, 5, **10**, 20 |

| | | | | | the region (used if the region's action type is border) | |
|---|---|---|---|---|---|---|
| `borderColor` | `couleurBordure` | `rahmenfarbe` | `kolorRamki` | no | color of border around the region (used if the region's action type is border) | Black<br>**Red**<br>#FF00FF<br>#00FF00FF |
| `nameOfTargetScene` | `nomDeScèneCible` | `nameDerZielszene` | `nazwaDocelowejSceny` | no | identifier of scene, to which the transition will be performed (used if the region's action type is transitionToScene) | scene02 |
| `nameOfImage` | `nomDuImage` | `bildname` | `nazwaRysunku` | no | identifier of image displayed in region | img01 |
| `nameOfSound` | `nomDuSon` | `tonname` | `nazwaDźwięku` | no | identifier of sound played | snd01 |
| `path` | `chemin` | `pfad` | `ścieżka` | no | separated by semicolon list of X,Y pairs (separated with comma) of integers describing the series of translations (used if the region's action type is move) | 0,-400;500,0 |
| `speed` | `vitesse` | `geschwindigkeit` | `szybkość` | no | speed of region translation given in pixels in seconds (used if the region's action type is move) | **0**, 1, 10, 100, 200 |
| `animationType` | `typeAnimation` | `animationstyp` | `typAnimacji` | no | one of six animation of region: no animation, enlarging and shrinking, rotation or swinging; the | en: **none** \| sizeChanging \| rotationCounterclockwise \| |
| | | | | | speed of the animation is determined by `animationPeriod` attribute which set the time the whole animation cycle takes place | rotationClockwise \| swingingHorizontal \| swingingVertical<br>fr: **aucun** \| changementTaille \| rotationSensInverseAiguilleMontre \| rotationSensAiguilleMontre \| oscillationHorizontal \| oscillationVerticale<br>de: **keine** \| größenänderung \| nachLinksDrehen \| nachRechtsDrehen \| horizontalSchwingen \| verticalSchwingen<br>pl: **brak** \| powiększanie \| obrotyWLewo \| obrotyWPrawo \| kołysanieWPoziomie \| kołysanieWPionie |
| `animationAmplitude` | `amplitudeAnimation` | `animationsamplitude` | `amplitudaAnimacji` | no | amplitude of animation; it can be relative to the size of region or given straightly in pixels | **-1**, 1, 10, 100 |
| `animationPeriod` | `périodeAnimation` | `animationsdauer` | `okresAnimacji` | no | period of animation (in miliseconds) | **-1**, 1, 10, 100 |
| `tag` | `étiquette` | `tag` | `znacznik` | no | used in case of SMI eyetrackers; string sending | *any string* |
| | | | | | to eyetracker server as a message marking the region's state enter | |
| `delayedTag` | `étiquetteRetard` | `verschobeneTag` | `znacznikOpóźniony` | no | as above, but sent with delay given by `delayOfDelayedTag` attribute | *any string* |
| `delayOfDelayedTag` | `retardateurEtiquetteRetard` | `verschiebungDerVerschobeneTag` | `opóźnienieZnacznikaOpóźnionego` | no | number of miliseconds after which delayed tag is sent | **-1**, 1, 10, 100 |
| `text` | `texte` | `text` | `tekst` | no | text of the inscription | *empty string* |
| `font` | `police` | `schrift` | `czcionka` | no | font of the inscription | **Times New Roman** <br> Arial <br> Courier New |
| `fontSize` | `taillePolice` | `schriftgröße` | `rozmiarCzcionki` | no | text size of the inscription | 10, **15**, 30 |
| `fontStyle` | `stylePolice` | `schriftart` | `stylCzcionki` | no | text decoration used for the inscription | string containing letters: <br> i – italic <br> b - bold <br> u – underline <br> s – strikeout <br> eg. ibus for all text decorations, default is empty string (no decoration) |
| `fontColor` | `couleurPolice` | `schriftfarbe` | `kolorCzcionki` | no | font color of the inscription | White <br> **Black** <br> #FF00FF <br> #00FF00FF |
| turnOffWhenFinished | éteindreQuandTerminé | nachAbschlussAusschalten | wyłączPoZakończeniu | no | disable current region after the state action is finished | en: yes \| **no**<br>fr: oui \| **non**<br>de: ja \| **neine**<br>pl: tak \| **nie** |
| nameOfRegionEnabledWhenStarted | nomDeRégionActivéeQuandDémarrée | nameDesBereichsNachAnfangEingeschaltet | nazwaObszaruWłączanegoPoRozpoczęciu | no | enable pointed region after the state action is started | rgn05 |
| nameOfRegionDisabledWhenStarted | nomDeRégionDésactivéQuandDémarrée | nameDesBereichsNachAnfangAusgeschaltet | nazwaObszaruWyłączanegoPoRozpoczęciu | no | disable pointed region after the state action is started | rgn06 |
| nameOfRegionEnabledWhenFinished | nomDeRégionActivéeQuandTerminée | nameDesBereichsAmEndeEingeschaltet | nazwaObszaruWłączanegoPoZakończeniu | no | enable pointed region after the state action is finished | rgn07 |
| nameOfRegionDisabledWhenFinished | nomDeRégionDésactivéQuandTerminée | nameDesBereichsAmEndeAusgeschaltet | nazwaObszaruWyłączanegoPoZakończeniu | no | disable pointed region after the state action is finished | rgb08 |

Tab. A.7. Attributes of `region` element, which cannot be overrided in `activation` and `reaction` sub-elements

| region / région / region / obszar (one can use rand prefix and % suffix) | | | | | | |
|---|---|---|---|---|---|---|
| English | French | Deutsch | Polish | Req. | Interpretation | Default (bold) and sample values |
| **`name`** | **`nom`** | **`name`** | **`nazwa`** | yes | identifier | rgn01 |
| `enabled` | `activé` | `aktiv` | `włączony` | no | region is enabled and visible after application starts | |
| **`locationOfCenterX`** **`locationOfCenterY`** | **`localisationDuCentreX`** **`localisationDuCentreY`** | **`positionDesMittelpunktsX`** **`positionDesMittelpunktsY`** | **`położenieŚrodkaX`** **`położenieŚrodkaY`** | yes | *x* and *y* coordinates of region's center position | *integer number* |
| **`sizeX`** **`sizeY`** | **`tailleX`** **`tailleY`** | **`größeX`** **`größeY`** | **`rozmiarX`** **`rozmiarY`** | yes | width and height of region | *integer number* |
| `shape` | `format` | `form` | `kształt` | no | determines the shape of region | pl: **prostokąt** \| koło \| elipsa |
| `offsetOfImageCenterX` `offsetOfImageCenterY` | `offsetDeLImageCentreX` `offsetDeLImageCentreY` | `abstandDesMittelpunktsDerZeichnungX` `abstandDesMittelpunktsDerZeichnungY` | `przesunięcieŚrodkaRysunkuX` `przesunięcieŚrodkaRysunkuY` | no | *x* and *y* coordinates of image's center offset relative to the region's center location | **0**, 10, 100 |
| `imageSizeX` `imageSizeY` | `tailleImageX` `tailleImageY` | `zeichnungsGrößeX` `zeichnungsGrößeY` | `rozmiarRysunkuX` `rozmiarRysunkuY` | no | width and height of the image displayed in region | *integer number; by default size of image is equal to size of region* |
| offsetOfTextX offsetOfTextY | offsetDuTexteX offsetDuTexteY | textAbstandX textAbstandY | przesunięcieTekstuX przesunięcieTekstuY | no | *x* and *y* coordinates of text location displayed in region relative to the region's location | **0**, 10, 100 |
| offsetOfActivationBarX offsetOfActivationBarY | offsetDuBarreDActivationX offsetDuBarreDActivationY | aktivierungsbalkenAbstandX aktivierungsbalkenAbstandY | przesunięciePaskaAktywacjiX przesunięciePaskaAktywacjiY | no | *x* and *y* coordinates of activation bar relative to its default position (under bottom left corner of region) | **0**, 10, 100 |
| regionAnimationEnabled | animationDeRégionActivé | animationDerRegionAktiv | animacjaObszaru | no | region animation is enabled (animation defined for current state is used) | en: **yes** \| no fr: **oui** \| non de: **ja** \| neine pl: **tak** \| nie |
| imageAnimationEnabled | animationImageActivé | zeichnungsAnimation | animacjaRysunku | no | image animation (independent of region animation) is enabled (animation defined for current state is used) | en: **yes** \| no fr: **oui** \| non de: **ja** \| neine pl: **tak** \| nie |
| conditionOfReactionCompletion | conditionDeRéalisationDeRéaction | bedingungDesEndesDerReaktion | warunekZakończeniaReakcji | no | condition of finishing the reaction state of the region and returning to normal state | en: **regionLeave** \| soundEnding \| timeElapsed fr: **quitterRégion** \| finSon \| tempsEcoulé de: **bereichVerlassen** \| klangStoppen \| verstricheneZeit pl: **wyjścieZObszaru** \| zakończenieOdtwarzaniaDźwięku \| upłynięcieCzasu |
| reactionDuration | duréeRéaction | reaktionsdauer | czasUtrzymywaniaReakcji | no | the time the region remains in reaction state (if the condition `timeElapsed` is choosen in `conditionOfReactionCompletion` attribute) | 10, 100, **-1** |
| holdSceneTransitionUntillReactionCompleted | garderTransitionSceneJusquAFinRéaction | verschiebDenÜbergangZuSceneBisZumEndeDerReaktion | odłóżPrzejścieDoScenyDoZakończeniaReakcji | no | holds the scene transition initiated by region until it returns to normal state | en: yes \| **no**<br>fr: oui \| **non**<br>de: ja \| **neine**<br>pl: tak \| **nie** |
| automaticReactionAfterTime | réactionAutomatiqueAprèsTemps | automatischeReaktionNachDerZeit | automatycznaReakcjaPoCzasie | no | time from entering to scene, after which the region switch to reaction state automatically | 100, 1000, **-1** *(by default this feature is switched off)* |
| ableToActivateBlackout | capableDActiverBlackout | imstandeZuAktivierenDenBlackout | możeAktywowaćZaczernienie | no | after region is in reaction state, the rest of scene is blackout (see the `scene` element's attributes relative to blackout) | en: **yes** \| no<br>fr: **oui** \| non<br>de: **ja** \| neine<br>pl: **tak** \| nie |
| resetAfterEnabled | réinitialiserAprèsActivation | nachEinschaltenZurücksetzen | resetujPoWłączeniu | no | the region is reset each time it is disabled and enabled again | en: yes \| **no**<br>fr: oui \| **non**<br>de: ja \| **neine**<br>pl: tak \| **nie** |
| ignoreGaze | ignorerAperçu | blickIgnorieren | ignorujSpojrzenie | no | region does not switch the states in reaction to gaze | en: yes \| **no**<br>fr: oui \| **non**<br>de: ja \| **neine**<br>pl: tak \| **nie** |
| enablingDelay | activationRetardement | einschaltenVerschieben | opóźnienieWłączenia | no | the delay after which the region is enabled (e.g. by | 0, 10, 100, **-1** |

| | | | | | other region's action) | |
|---|---|---|---|---|---|---|
| `disablingDelay` | désactivationRetardement | ausschaltenVerschieben | opóźnienieWyłączenia | no | the delay after which the region is disabled (e.g. by other region's action or using attribute `turnOffWhenFinished`) | 0, 10, 100, **-1** |
| `reactionKey` | cléRéaction | reaktionTaste | klawiszReakcji | no | keyboard key which switch the region to reaction state | A, B, C, ... |
| `onActivationCompleted` | surActivationComplétée | nachAktivierung | poAktywacji | no | name of method from .NET library, which will be executed after activation of region | methodScene01Rgn01Activation |
| `onReactionStarted` | surRéactionDémarée | nachAnfangDerReaktion | poRozpoczęciuReakcji | no | as above, but for beginning of the reaction state | methodScene01Rgn01ReactionStarted |
| `onReactionFinished` | surRéactionComplétée | nachEndeDerReaktion | poZakończeniuReakcji | no | as above, but for finishing of the reaction state | methodScene01Rgn01ReactionFinished |
| `onNormalStateReturn` | surRetourEtatNormal | nachNormalzustandZurücksetzen | poPowrocieDoStanuNormalnego | no | as above, but for returning to normal state | methodScene01Rgn01ReturnToNormalState |
| `onStateChanged` | surChangementDEtat | amZustandWechseln | poZmianieStanu | no | as above, but for any state change | methodScene01Rgn01StateChanged |